%% file: HIN-21-012_temp.tex
\pdfoutput=1
\documentclass[11pt,twoside,a4paper,cmspaper,final,collab]{cms-tdr}
\def\svnVersion{c8f389e}\def\svnDate{2024/10/06}\def\cmsCernNoTag{CERN-EP-2024-011}\def\cmsCernDate{\today}\def\cmsMessage{Submitted to Physical Review Letters}
\begin{document}\cmsNoteHeader{HIN-21-012}

\newlength\cmsFigWidth
\ifthenelse{\boolean{cms@external}}{\setlength\cmsFigWidth{0.98\columnwidth}}{\setlength\cmsFigWidth{0.8\textwidth}}
\ifthenelse{\boolean{cms@external}}{\providecommand{\cmsLeft}{top\xspace}}{\providecommand{\cmsLeft}{left\xspace}}
\ifthenelse{\boolean{cms@external}}{\providecommand{\cmsRight}{bottom\xspace}}{\providecommand{\cmsRight}{right\xspace}}

\ifthenelse{\boolean{cms@external}}{\renewcommand{\appendixname}{Endmatter}}{}

\newcommand {\roots}    {\ensuremath{\sqrt{s}}}
\newcommand {\pp}  {\ensuremath{\text{pp}}\xspace}
\newcommand {\pPb}  {\ensuremath{\text{pPb}}\xspace}
\newcommand {\pAu}  {\ensuremath{\text{pAu}}\xspace}
\newcommand {\dAu}  {\ensuremath{\text{dAu}}\xspace}
\newcommand {\AuAu}  {\ensuremath{\text{AuAu}}\xspace}
\newcommand {\CuCu}  {\ensuremath{\text{CuCu}}\xspace}
\newcommand {\PbPb}  {\ensuremath{\text{PbPb}}\xspace}
\newcommand{\nch}    {\ensuremath{N_\text{ch}}\xspace}
\newcommand{\mpt}{\ensuremath{[p_{\mathrm{T}}]}\xspace}
\newcommand{\noff}    {\ensuremath{N_\mathrm{trk}^\mathrm{offline}}\xspace}
\newcommand{\nchrec}    {\ensuremath{N_\mathrm{ch}^\mathrm{rec}}\xspace}
\newcommand{\llangle}{\ensuremath{\langle\langle}}
\newcommand{\rrangle}{\ensuremath{\rangle\rangle}}
\newcommand{\PYTHIAviii}{\textsc{pythia}8\xspace}

\cmsNoteHeader{HIN-21-012}
\title{Correlations between azimuthal anisotropy and mean transverse momentum in \texorpdfstring{\pp}{pp}, \texorpdfstring{\pPb}{pPb}, and peripheral \texorpdfstring{\PbPb}{PbPb} collisions}

\date{\today}

\abstract{
Correlations between azimuthal anisotropy and mean transverse
momentum of charged particles in proton-proton (\pp), proton-lead (\pPb), and 
peripheral lead-lead (\PbPb) collisions are presented
as a function of charged particle multiplicity. 
The \pp, \pPb and \PbPb collision data were
collected using the CMS detector at the LHC
with a center-of-mass energy per nucleon pair of
13, 8.16 and 5.02\TeV, respectively.
The two- and four-particle cumulants for
the second- and third-order Fourier anisotropy harmonics are correlated
with the mean transverse momentum of charged particles on an event-by-event basis. 
In \pp and \pPb systems, the observed correlation coefficients 
based on two-particle cumulants are found to change 
from negative to positive values as the charged particle 
multiplicity decreases. The sign changes disappear 
when the correlated particles are required to be 
further apart in pseudorapidity. 
Additionally, no sign changes in correlation coefficients are 
observed when employing four-particle cumulants.
Models incorporating initial-state gluon saturation and 
final-state hydrodynamic evolutions are compared to 
\pPb data and the predicted sign changes are not observed.
}

\hypersetup{
pdfauthor={CMS Collaboration},
pdftitle={Correlations between azimuthal anisotropy and mean transverse momentum in \pp, \pPb, and peripheral \PbPb collisions},
pdfsubject={CMS},
pdfkeywords={CMS, physics, heavy ion, flow, mean transverse momentum}}

\maketitle 

A hot and dense medium known as the 
quark-gluon plasma (QGP) 
has been extensively studied using heavy ion
collisions at the BNL RHIC~\cite{Alver:2008aa,
Adams:2005ph,Abelev:2009af,Alver:2009id} 
and CERN LHC~\cite{Chatrchyan:2011eka,ALICE:2011ab,
Aamodt:2011by,Chatrchyan:2012wg,Aamodt:2010pa,ATLAS:2012at,Chatrchyan:2012ta, CMS:2024krd}. 
The azimuthal anisotropy of the produced particles in these collisions is a 
powerful tool to study the collective dynamics and transport 
properties of the QGP. This anisotropy is characterized by 
the Fourier coefficients ($v_n$) of the particle azimuthal angle ($\phi$) 
distribution $\ddinline{N}{\phi} \propto 1 + 2 \sum_{n} v_{n} \cos[n(\phi - \Psi_{n} )]$, 
where $\Psi_{n}$ represents the phase of the $n^{th}$-order 
azimuthal flow vector (also referred to as the angle of 
the $n^{th}$-order symmetry plane). 
In a hydrodynamic picture, the Fourier anisotropy harmonics 
are the result of strong partonic rescatterings in the final 
state, responding to the initial geometry of the 
colliding system~\cite{Alver:2010dn,Schenke:2010rr,Qiu:2011hf}.
In the past decade, a remarkable similarity in the azimuthal 
anisotropy signatures has been observed between heavy ion
collisions and smaller collision systems,
such as proton-proton (\pp), proton-lead (\pPb), and proton-gold (\pAu)~\cite{Khachatryan:2010gv,Aad:2015gqa,Khachatryan:2015lva, CMS:2012qk,Abelev:2012ola,Aad:2012gla,Aaij:2015qcq, Aidala:2017ajz,Aidala:2016vgl,Adare:2015ctn,PHENIX:2018lia}.
The similarity holds even for multiparticle correlations, 
which can suppress ``nonflow" effects that do not result 
from correlations related to the bulk properties of the 
medium~\cite{Aidala:2017ajz, CMS:2013jlh}.  These effects can include back-to-back jet 
correlations and resonance 
decays, which usually produce particles in localized 
pseudorapidity ($\eta$) regions~\cite{CMS:2015yux, CMS:2016fnw, CMS:2019wiy}. 
Apart from the final-state effects, the observed anisotropy in small 
systems can also originate from initial-state effects, for example, 
in the color-glass condensate (CGC)
effective theory~\cite{Dusling:2012cg, Dumitru:2014vka, Schenke:2015aqa}.
The dominant origin of the azimuthal 
anisotropy in small systems is still under active 
discussion~\cite{Nagle:2018nvi, Busza:2018rrf}, 
because no observable has been found to unambiguously 
distinguish between the final- and initial-state effects.

In addition to generating the final-state azimuthal anisotropy 
the hydrodynamic response 
also results in a radial flow,
which contributes to the mean 
transverse momentum $[\pt]$ on an event-by-event basis. 
The correlations between radial and anisotropic flow can be 
quantified using a modified linear correlator~\cite{Mazeliauskas:2015efa, Bozek:2016yoj},
\begin{linenomath}
\begin{equation}
\label{eq:corr_intro}
  \rho(v_n^{2} , \mpt ) = \frac{\cov(v_n^{2} , \mpt )}{\sqrt{\mathrm{Var}(v_n^{2})_\text{dyn}}\sqrt{\mathrm{Var}( \mpt)_\text{dyn}}},
\end{equation}
\end{linenomath}
where $\cov(v_n^{2} , \mpt )$ is the covariance 
between $v_{n}^2$ and \mpt, 
and Var$(v_{n}^2)_\text{dyn}$ and Var$(\mpt)_\text{dyn}$ are the dynamical variances of 
the $v_{n}^2$ and \mpt distributions, respectively. 
Dynamical variances remove auto-correlation effects, when compared
with variances of $v_{n}^2$ and \mpt distributions, and better capture intrinsic 
initial-state fluctuations.
This correlator defined by Eq.~(\ref{eq:corr_intro}) is sensitive to the 
details of the initial conditions, and its magnitude can be traced 
back to the initial density profile of the nuclear 
overlap~\cite{Schenke:2020uqq, Giacalone:2020dln}. 

Recently it was suggested that this correlator 
might be able to distinguish between initial- 
and final-state effects~\cite{Giacalone:2020byk}. 
Specifically, the CGC model predicts a sign change 
of the $\rho(v_n^{2},\mpt)$ correlator  
going from high-multiplicity (dominated by 
final-state effects) to very low-multiplicity (dominated by 
initial-state effects) events, in small collision systems.
A prior ATLAS measurement using two-particle cumulants 
shows no sign change across the explored multiplicity 
range in \pPb collisions~\cite{ATLAS:2019pvn}.
However, these results do not cover the multiplciity 
range where the initial-state sign change is expected. 
It was realized that a sign change could exist due to
nonflow effects as shown by the \PYTHIAviii event generator
with the A2 tune~\cite{Zhang:2021phk, Lim:2021auv}.
Measurements of the correlator with proper treatment of 
nonflow effects, and the searches for 
correlator sign changes in low-multiplicity \pp, \pPb, 
and peripheral \PbPb collisions can provide insights into 
the origin of the azimuthal correlations in small systems.

In this Letter, these correlations are measured to very 
low multiplicity regions in three collision systems where 
sign changes are predicted due to the dominance of 
initial-state effects. Measurements using four-particle 
cumulants are also presented for the first time to further 
suppress nonflow effects.
The results in pPb collisions are compared directly with CGC 
predictions.
In addition, the correlators for the third Fourier 
harmonic are also presented as a function of charged particle multiplicity.
Tabulated results are provided in the HEPData record for this analysis~\cite{hepdata}.

The CMS apparatus~\cite{CMS:2008xjf,CMS:2309.05466} is a multipurpose, nearly hermetic detector, designed to trigger on and identify electrons, muons, photons, and (charged and neutral) hadrons~\cite{CMS:2020uim,CMS:2018rym,CMS:2014pgm}. A global ``particle-flow'' (PF) algorithm~\cite{CMS:2017yfk} aims to reconstruct all individual particles in an event, combining information provided by the all-silicon inner tracker and by the crystal electromagnetic and brass-scintillator hadron calorimeters, operating inside a 3.8\unit{T} superconducting solenoid, with data from the gas-ionization muon detectors embedded in the flux-return yoke outside the solenoid. 
Forward hadronic calorimeters (HF), made of steel and quartz fibers, extend the $\eta$ 
coverage provided by the barrel and endcap detectors up to $\abs{\eta} \sim 5$.
The silicon tracker used in 2016 measured charged particles within the range $\abs{\eta} < 2.5$. For nonisolated particles of $1 < \pt < 10\GeV$ and $\abs{\eta} < 1.4$, the track resolutions were typically 1.5\% in \pt and 25--90 (45--150)\mum in the transverse (longitudinal) impact parameter~\cite{CMS:2014pgm}. At the start of 2017, a new pixel detector was installed~\cite{Phase1Pixel}; the upgraded tracker measured particles up to $\abs{\eta} < 3.0$ with typical resolutions of 1.5\% in \pt and 20--75\mum in the transverse impact parameter~\cite{DP-2020-049} for nonisolated particles of $1 < \pt < 10\GeV$.
The data sample is collected with a two-level trigger system: 
at level-1 events are selected by custom hardware processors and  
the high-level trigger uses fast versions of the offline software~\cite{CMS:2020cmk,CMS:2016ngn}.

The measurements presented in this Letter 
use \pp, \pPb and \PbPb collisions with center-of-mass energy 
per nucleon pairs of 13, 8.16 and 5.02\TeV, taken 
in 2018, 2016, and 2018, and with 
integrated luminosities of 28.6\pbinv,
186\nbinv and 0.607\nbinv, respectively~\cite{CMS:LUM-18-002,CMS:LUM-17-002,CMS:LUM-18-001,CMS:LUM-17-003}. 
When measuring the correlations in \pp and \pPb collisions, 
the same beam crossing may contain multiple independent interactions, 
which constitute a background for the analysis of high-multiplicity 
events. The average number of collisions per bunch
crossing (pileup) in \pp (\pPb) data varied between 0.1 and 1.3 (0.10 and 0.25). 
A procedure similar to that described 
in Ref.~\cite{CMS:2013jlh} is used for identifying and rejecting pileup.
Minimum bias (MB) \pp and \pPb events are triggered by energy deposits in at least one of
the two HFs above a threshold 
of approximately 1\GeV, and the presence of at least
one track with $\pt > 0.4\GeV$ in the pixel tracker for \pPb collisions.
Events are also required
to contain a primary vertex within 15\unit{cm} of the nominal 
interaction point along the beam axis
and 0.2\unit{cm} in the transverse direction. 
The primary vertex is chosen as the reconstructed vertex with the largest number of associated tracks
At least two reconstructed tracks were required to be associated with the primary vertex.
The MB \PbPb events are selected to have signals above readout thresholds 
in the range of ${\sim}$6--12\GeV on both sides of the HF calorimeters.
The \PbPb events are further filtered to have a primary vertex within 15\unit{cm} of the nominal
interaction point along the beam axis
and 0.2\unit{cm} in the transverse direction, and at least two towers with total deposited energy 
larger or equal to 4\GeV in each of the HF detectors.
The trigger, event reconstruction, and selections 
are described in previous correlation analyses~\cite{CMS:2013jlh, CMS:2015yux, CMS:2016fnw, CMS:2019wiy}.
For all data sets analyzed, primary tracks, i.e., tracks that originate from the primary vertex and
satisfy the high-purity criteria of Ref.~\cite{CMS:2014pgm},   
are used to perform the correlation measurements.
In addition, the impact parameter significance of the tracks with 
respect to the primary vertex in
the longitudinal and transverse direction is required to be less than 3 standard deviations.
The relative \pt uncertainty must be less than 10\% for the \pt range used. 
To ensure high tracking efficiency and minimal background contamination only
tracks with $\abs{\eta} < 2.4$ and $\pt > 0.5\GeV$ are used, as in Ref.~\cite{CMS:2014pgm}.
The selected tracks are corrected for tracking inefficiency and acceptance 
found using simulated Monte Carlo
samples from \PYTHIA 8.212~\cite{Sjostrand:2014zea} tune CP5~\cite{Sirunyan:2019dfx},
\HIJING v1.35~\cite{Gyulassy:1994ew}, and \HYDJET 1.9~\cite{Lokhtin:2005px} for \pp, \pPb,
and \PbPb, respectively.
If not specified otherwise, \PYTHIAviii refers to this CP5 tune throughout the paper.
The response of the CMS detector to these simulated events is based
on \GEANTfour~\cite{GEANT4}.

All previous studies~\cite{ATLAS:2019pvn, ALICE:2021gxt, Giacalone:2020byk, Zhang:2021phk, Lim:2021auv} of 
the modified linear correlator use $v_{n}^2$ from two-particle 
correlations in Eq.~(\ref{eq:corr_intro}).
The $v_n^2$ term can be found from 
the Q-cumulant method described in Ref.~\cite{Bilandzic:2010jr}.
The two- and four-particle $v_n$ in this method can be written as,
\begin{linenomath}
\begin{equation}
  v_n\{2\} = \sqrt{c_n\{2\}}, \quad
  v_n\{4\} = \sqrt[4]{-c_n\{4\}}. 
  \label{eqn:vn_2p_4p}
\end{equation}
\end{linenomath}
The two- and four-particle cumulants $c_n\{2\}$ and $c_n\{4\}$ are
\begin{linenomath}
\begin{equation}
  c_n\{2\} = \llangle {2} \rrangle, \quad
  c_n\{4\} = \llangle {4} \rrangle - 2 \langle \langle 2 \rangle \rangle^2, 
  \label{eqn:cn_2p_4p}
\end{equation}
\end{linenomath}
where the two- and four-particle 
correlations $\langle \langle 2 \rangle \rangle$ and $\langle \langle 4 \rangle  \rangle$ 
are from
\begin{linenomath}
\begin{equation}
  \llangle {2} \rrangle = \llangle \re^{in(\phi_{1}-\phi_{2})} \rrangle,\quad
  \llangle {4} \rrangle = \llangle \re^{in(\phi_{1}+\phi_{2}-\phi_{3}-\phi_{4})} \rrangle.
  \label{eqn:corr_2p_4p}
\end{equation}
\end{linenomath}
Here, $\phi_{\mathrm{j}}$ (j = 1, \ldots, 4) are the azimuthal angles of four 
different particles in an event, and the double average
symbol $\llangle\cdots\rrangle$ indicates that the average is 
taken over all particles from all events.

To suppress nonflow effects in the $v_n\{2\}$, all previous 
measurements applied the ``subevent method", which 
separates each event into different subsets in   
separated $\eta$ ranges~\cite{ATLAS:2019pvn, ALICE:2021gxt}. 
With two subevents for $c_n\{2\}$, the covariance of the correlator 
in Eq.~(\ref{eq:corr_intro}) is
\begin{linenomath}
\begin{equation}
    \cov(c_n\{2\},\mpt) = \Re{\Big\langle \sum_{a,b}\exp^{in(\phi_a-\phi_b)}\big(\mpt - \langle\mpt\rangle\big)\Big\rangle},
\end{equation}
\end{linenomath}
where $\phi_a$ and $\phi_b$ are the azimuthal angles of 
particles $a$ and $b$ in subevents A and B, respectively.
The $\langle\mpt\rangle$ is the average \mpt in all the
events in a certain multiplicity range.
We select tracks with $-2.4 < \eta < -0.75$ to be subevent A, 
whereas tracks 
with $0.75 < \eta < 2.4$ belong to subevent B. Tracks from the
third subevent in the central region 
$\abs{\eta} < 0.5$ are used to obtain \mpt in each event.
These selections ensure that subevents A and B are symmetric in $\eta$, and
that there is an $\eta$ gap between each subevent to reduce nonflow effects. 

The dynamical variance of $c_n\{2\}$ and \mpt are derived 
in Ref.~\cite{Bozek:2016yoj}:
\begin{linenomath}
\begin{equation}
    \mathrm{Var}(c_n\{2\})_{\text{dyn}} = \langle \langle 4 \rangle \rangle - \langle \langle 2 \rangle \rangle^2,
\end{equation}
\end{linenomath}
\begin{linenomath}
\begin{equation}
    \mathrm{Var}( \mpt)_\text{dyn} = \Big\langle \big[ (p_{\textrm{T}i}-\langle\mpt\rangle)(p_{\textrm{T}j}-\langle\mpt\rangle) \big] \Big\rangle, 
\end{equation}
\end{linenomath}
where the indices $i \neq j$ and both run over all charged particles within $\abs{\eta} < 0.5$.

The correlator $\rho(c_n\{2\},\mpt)$ is extracted using 
particles in the range $0.5 < \pt <5.0\GeV$. Particles with
$\pt > 5\GeV$ are excluded to reduce nonflow effects. The analysis
is repeated in different ranges of the number of reconstructed 
charged particles \nchrec, which is obtained with the same 
\pt range as the correlator within $\abs{\eta}<2.4$. 
The results are presented as a function of \nch, which is found 
by applying acceptance and efficiency corrections to \nchrec. 
For each \nchrec range used in this analysis, 
Table~\ref{tab:nch_ntrkoffline} of the \appendixname\ reports the mean values of \nch and also of \noff, the 
number of offline-reconstructed charged particles with
$\pt >0.4\GeV$ and $\abs{\eta}<2.4$ used in previous
CMS measurements~\cite{Khachatryan:2010gv, CMS:2012qk, CMS:2013jlh}.
Since we are interested in small collision 
systems, only peripheral events with $\nch<400$ are 
presented for \PbPb collisions.

Studies using A2 tune of \PYTHIAviii~\cite{Zhang:2021phk, Lim:2021auv} show 
that the remaining nonflow contributions to $c_2\{2\}$ are not 
negligible when going to very low-multiplicity regions, 
even with the subevent methods. 
In these studies, we improve the suppression of nonflow effects 
with two approaches.
In the first approach, we increase the minimum $\eta$ gap between subevents A and B 
from 1.5 to 2.0, by changing the $c_2\{2\}$ analysis using particles 
from $\abs{\eta}>0.75$ to $\abs{\eta}>1.00$. Nonflow correlations between
particles inside back-to-back jets in the forward $\eta$ regions still 
remain with this larger $\eta$ gap~\cite{CMS:2013jlh}.

In the second approach, we extend the current observable by
replacing $c_2\{2\}$ with four-particle cumulant $c_2\{4\}$. 
Although nonflow contributions to multiparticle cumulants are 
less dominant when compared with 
two-particle correlations,
using the subevent method is important to remove remaining
nonflow contributions in multiparticle correlations~\cite{Jia:2017hbm}.
Particles with $0.75 < \abs{\eta} < 2.4$ are divided into 
three equal sized $\eta$ intervals to 
obtain $c_2\{4\}$ for each event. 
These intervals are $-2.4 < \eta < -1.3$, $0.75 < \abs{\eta} < 1.3$,
and $1.3 < \eta < 2.4$.
The event-by-event
$c_2\{4\}$ is then correlated with \mpt in the same event. 
When nonflow contributions are not negligible, the
method using $c_2\{4\}$ suppresses them much more than
the $c_2\{2\}$ method. The correlator
using $c_2\{4\}$ is more sensitive to the presence of a 
sign change in the data.
Corrections for tracking efficiency and misreconstructed 
tracks are done at particle level for both the cumulants and \mpt.

The systematic uncertainties in the experimental 
procedure are evaluated as a function of \nch
by using alternative procedures for extracting
the correlators.
Systematic uncertainties due to tracking inefficiency and misreconstructed
track rate are studied by varying the track quality requirements. 
The selection thresholds on the significances of the transverse 
and longitudinal track impact parameters 
divided by their uncertainties are varied from 2 to 5.  
In addition, the upper limit on the relative \pt uncertainty
is varied from 5 to 10\%. 
The resulting systematic uncertainty 
is from
$2\times10^{-7}$ to $5\times10^{-6}$ for the 
covariances $\cov(c_n\{2\}, \mpt)$ and $\cov(c_n\{4\}, \mpt)$, 
and from 0.001 to 0.008
for the correlator $\rho(c_{n}\{2\}, \mpt)$, depending 
on multiplicity and collision systems.
The sensitivity of the results to the
primary vertex position along the beam axis ($z_\text{vtx}$) is 
quantified by comparing events with different $z_\text{vtx}$
locations from $-15$ to $+15$\unit{cm}.
The magnitude of this uncertainty is estimated to
be from $1\times10^{-7}$ to $2\times10^{-5}$ for the
covariances, and from 0.001 to 0.018 for $\rho(c_{n}\{2\}, \mpt)$, 
depending on multiplicity and collision system. 
Systematic effects from event selections are explored
by comparing results obtained with and without 
the requirement of the coincidence of HF calorimeter tower
signals above the threshold. 
The pileup effect is studied by requiring the presence of
only a single reconstructed vertex.
Systematic uncertainties from event selections and pileup
are both found to be less than $6\times10^{-6}$ 
for the covariances, and less than 0.006 for
the correlators in all collision systems.
The tracking and vertex position selections are the dominant sources of uncertainty.
Systematic uncertainties originating from different 
sources are added in quadrature to obtain the total 
systematic uncertainty.

\begin{figure}[hbt!]
\centering
\includegraphics[width=\cmsFigWidth]{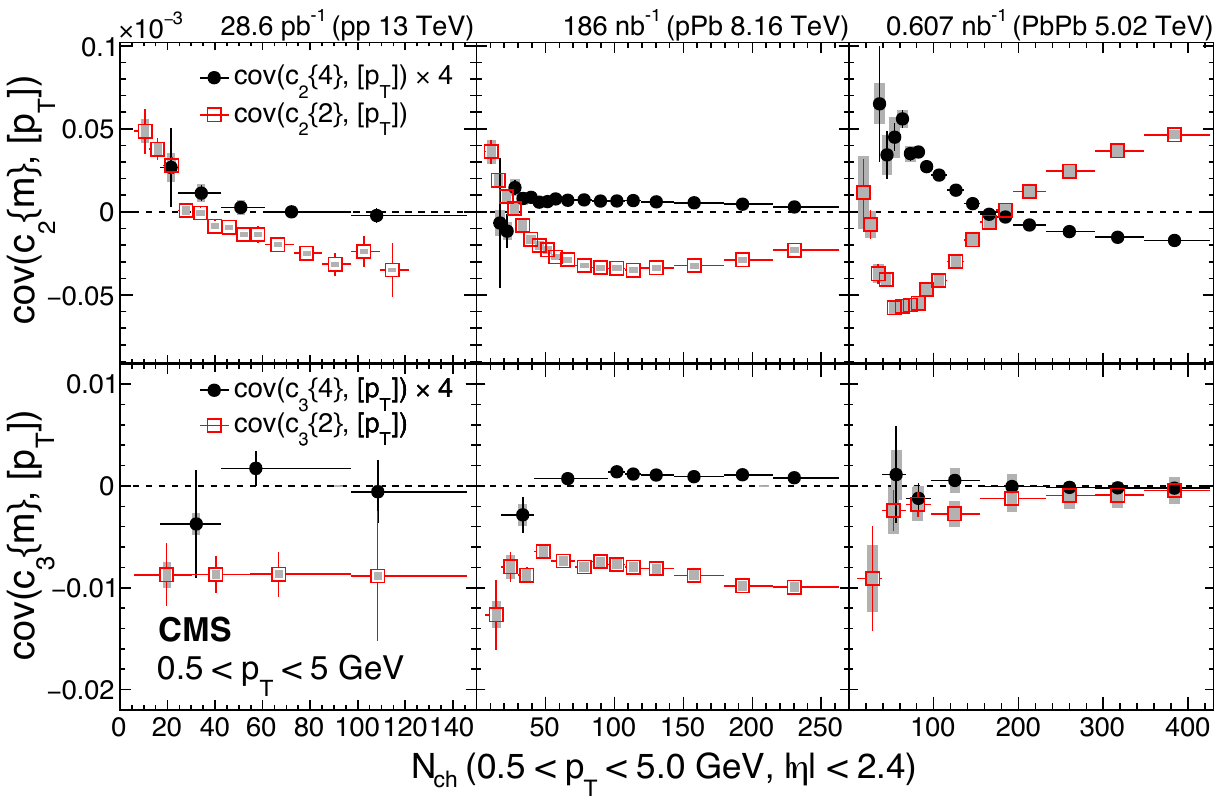}
  \caption{
   The covariances of cumulants from two- and four-particle correlations and \mpt as a     function of charged particle multiplicity (\nch) in 13\TeV  \pp (left), 8.16\TeV \pPb      (middle), and 5.02\TeV \PbPb (right) collisions. The two-particle cumulants are obtained with $\abs{\eta}>0.75$. The upper (lower) panels show the second  (third) harmonics. The error bars correspond to statistical uncertainties, while the       shaded areas denote the systematic uncertainties.
   }
    \label{fig:covnm_3panels}
\end{figure}

The measurements of covariances from two- and four-particle 
correlations for the second- and third-order Fourier harmonics in 13\TeV  \pp,
8.16\TeV \pPb, and 5.02\TeV \PbPb collisions are 
presented in Fig.~\ref{fig:covnm_3panels}.
To compare $\cov(c_2\{2\}, \mpt)$ and $\cov(c_2\{4\}, \mpt)$
in the same scale, the values of $\cov(c_2\{4\}, \mpt)$ are
multiplied by $4$ in all of the panels.
In both \pp and \pPb collisions, $\cov(c_2\{2\}, \mpt)$ for
$\abs{\eta}>0.75$ exhibits  
a sign change from positive to negative as \nch increases.
This trend is qualitatively consistent with the expectation of 
a sign change from the CGC model.
However, no clear sign change
is observed in \pp and \pPb collisions 
for $\cov(c_2\{4\}, \mpt)$, which mitigates
nonflow contributions compared to $\cov(c_2\{2\}, \mpt)$.
The $\cov(c_2\{4\}, \mpt)$ values are consistent with $0$ in \pp collisions  
with the current statistical precision. 
As \nch becomes smaller in \PbPb collisions, the values of $\cov(c_2\{2\}, \mpt)$
change from positive to negative below $\nch \approx 180$,  
reach a minimum at $\nch \approx 60$, and
then approach zero at the lowest \nch range.

One feature of the observable with $c_2\{4\}$ is that,
when nonflow contributions are not dominant,
the variables using $c_2\{4\}$ and $c_2\{2\}$ have 
opposite signs and trends.
This is because the relations between cumulants and $v_n$ are
$c_{2}\{2\} = (v_2\{2\})^2$
and $c_{2}\{4\} = -(v_2\{4\})^4$
from Eq.~(\ref{eqn:vn_2p_4p}).
As $v_n$ increases, both
$v_2\{2\}$ and $v_2\{4\}$ become larger if nonflow and 
flow fluctuations do not dominate the flow signal.
Therefore, the values increase for $c_{2}\{2\}$ but
decrease for $c_{2}\{4\}$ because of the minus sign.
This opposing trends for $\cov(c_2\{2\}, \mpt)$ 
and $\cov(c_2\{4\}, \mpt)$
are clear for high-multiplicity \PbPb results 
in Fig.~\ref{fig:covnm_3panels}.

Regarding the third harmonic, $\cov(c_3\{2\}, \mpt)$ is negative for the 
full \nch range in the three collision systems.
The signs of $\cov(c_3\{2\}, \mpt)$ and $\cov(c_3\{4\}, \mpt)$ are 
opposite in high-multiplicity \pPb events, similar 
to $\cov(c_2\{2\}, \mpt)$ and $\cov(c_2\{4\}, \mpt)$.
In both \pp and \PbPb collisions, the $\cov(c_3\{4\}, \mpt)$ values are
consistent with $0$ with the current statistical precision.

\begin{figure}[thb!]
\centering
\includegraphics[width=\cmsFigWidth]{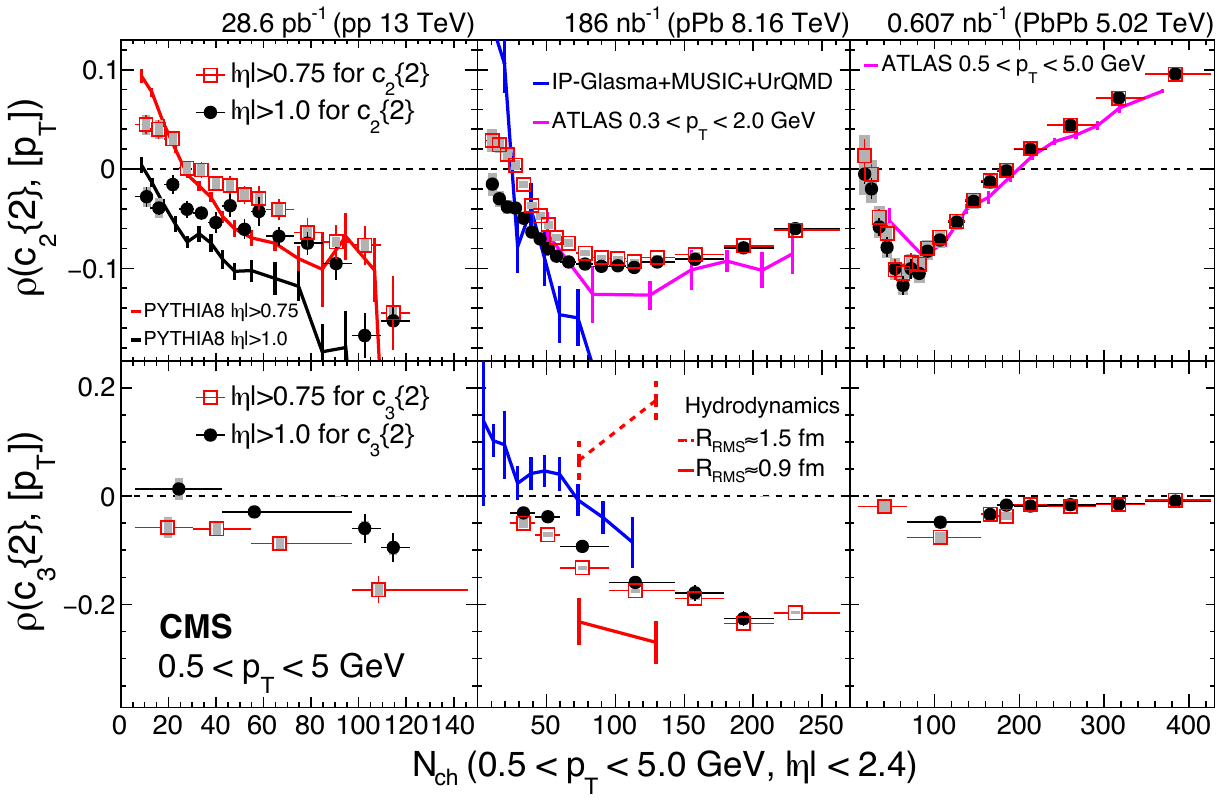}
  \caption{
   The correlator using two-particle cumulant from $\abs{\eta}>0.75$
and $\abs{\eta}>1.0$ as a function of \nch in 13\TeV \pp (left),
8.16\TeV \pPb (middle), and 5.02\TeV \PbPb (right) collisions.
The upper (lower) panels show the second (third) harmonics.
The error bars correspond to the statistical uncertainties,
and the shaded areas denote the systematic uncertainties.
Calculations from \PYTHIAviii (upper left panel, red and black lines) and
IP-Glasma+MUSIC+UrQMD (middle panels, blue lines)~\cite{Giacalone:2020byk}
are compared with the data.
Hydrodynamic predictions (lower middle panel, red lines)~\cite{Bozek:2020drh} with
an average RMS transverse radius of the initial fireball of either 0.9  or 1.5 fm are also included in the comparison.
The magenta lines in the middle and right upper panels 
represent the ATLAS results~\cite{ATLAS:2019pvn}
in \pPb and \PbPb collisions at 5.02\TeV using $\abs{\eta}>0.75$ with
the error bars denoting the statistical and systematic uncertainties added in quadrature.
   }
   \label{fig:rhon2_3panels}
\end{figure}

{\tolerance=800
The correlator with a larger $\eta$ gap ($\abs{\eta}>1.0$  corresponding to a minimum $\eta$ gap of 2.0) 
for the cumulants is shown
in Fig.~\ref{fig:rhon2_3panels}. In both \pp and \pPb collisions, 
the sign change at low \nch  
disappears with the larger $\eta$ gap between the two
subevents, which is also observed in calculations using \PYTHIAviii.  
The predictions in \pPb collisions at 5.02\TeV 
from the IP-Glasma+MUSIC+UrQMD model~\cite{Giacalone:2020byk} 
with $0.5 < \pt < 5.0\GeV$ are compared with the data in Fig.~\ref{fig:rhon2_3panels}. 
This model includes gluon saturation in the initial state followed by
hydrodynamic evolution and hadronic interactions.
The characteristic sign change of the
correlator predicted by this model is observed in data at the 
same \nch location for $\abs{\eta}>0.75$, but it 
disappears when using $\abs{\eta}>1.0$, which leads to less nonflow effects.
The results indicate that after removing more nonflow effects, the
CGC signal is not observed in the data.
More nonflow studies incorporated in the CGC 
model are needed to understand the origin of azimuthal anisotropy in small systems.
Results from the ATLAS Collaboration in \pPb (\PbPb) collisions at 5.02\TeV
with $0.3 < \pt < 2.0\GeV$ ($0.5 < \pt < 5.0\GeV$)~\cite{ATLAS:2019pvn} 
are also compared and found to be consistent with results obtained from this analysis.
\par}

The $n=3$ correlators measured with the larger $\eta$ gap are 
negative in all three collison systems.
The results are compared with predictions from a 
hydrodynamic simulation of pPb collisions~\cite{Bozek:2020drh} 
using $\pt > 0.5\GeV$ and having an average root-mean-square (RMS) transverse 
radius of the initial fireball of either 0.9 or 1.5 fm.
These sizes correspond to two versions of the Glauber model for 
the initial state~\cite{Bzdak:2013zma}: the smaller fireball 
with 0.9 fm deposits entropy in the nucleon overlap region between 
the participant nucleons, while the larger fireball with 1.5 fm 
deposits entropy at both positions of the participant nucleons. 
Qualitatively, the data are better described by the smaller 
initial fireball, suggesting that entropy deposition 
should occur in the nucleon overlap region rather
than at the center of both nucleons.
The predictions from \text{IP-Glasma+MUSIC+UrQMD} for $n=3$ also indicate a 
sign change, although it is not observed in data.

In summary, correlations between mean transverse 
momentum $[\pt]$ and multiparticle cumulants from two- and four-particle
correlations for Fourier harmonics $n=2$ and $n=3$ are presented 
in proton-proton (\pp) collisions at $\roots=13\TeV$, 
proton-lead (\pPb) collisions at $\sqrtsNN=8.16\TeV$ and 
peripheral lead-lead (\PbPb) collisions at $\sqrtsNN=5.02\TeV$.
Sign changes in the modified linear correlators 
are observed as a function of 
charged particle multiplicity when using
two-particle cumulants with a 
minimum $\eta$ gap of 1.5, in the \pp and \pPb systems. 
These sign changes disappear when nonflow effects are suppressed 
using a minimum $\eta$ gap of 2.0.
To further reduce nonflow contribution, four-particle 
cumulants $c_2\{4\}$ are also correlated with $[\pt]$ which shows 
no sign change in \pp and \pPb collisions, similar to the two-particle correlation results with a larger $\eta$ gap. 
This indicates that after removing nonflow effects, 
the sign change predicted at low multiplicity by 
initial-state saturation is not observed. 

\begin{acknowledgments}
  We congratulate our colleagues in the CERN accelerator departments for the excellent performance of the LHC and thank the technical and administrative staffs at CERN and at other CMS institutes for their contributions to the success of the CMS effort. In addition, we gratefully acknowledge the computing centers and personnel of the Worldwide LHC Computing Grid and other centers for delivering so effectively the computing infrastructure essential to our analyses. Finally, we acknowledge the enduring support for the construction and operation of the LHC, the CMS detector, and the supporting computing infrastructure provided by the following funding agencies: SC (Armenia), BMBWF and FWF (Austria); FNRS and FWO (Belgium); CNPq, CAPES, FAPERJ, FAPERGS, and FAPESP (Brazil); MES and BNSF (Bulgaria); CERN; CAS, MoST, and NSFC (China); MINCIENCIAS (Colombia); MSES and CSF (Croatia); RIF (Cyprus); SENESCYT (Ecuador); ERC PRG, RVTT3 and MoER TK202 (Estonia); Academy of Finland, MEC, and HIP (Finland); CEA and CNRS/IN2P3 (France); SRNSF (Georgia); BMBF, DFG, and HGF (Germany); GSRI (Greece); NKFIH (Hungary); DAE and DST (India); IPM (Iran); SFI (Ireland); INFN (Italy); MSIP and NRF (Republic of Korea); MES (Latvia); LMTLT (Lithuania); MOE and UM (Malaysia); BUAP, CINVESTAV, CONACYT, LNS, SEP, and UASLP-FAI (Mexico); MOS (Montenegro); MBIE (New Zealand); PAEC (Pakistan); MES and NSC (Poland); FCT (Portugal); MESTD (Serbia); MCIN/AEI and PCTI (Spain); MOSTR (Sri Lanka); Swiss Funding Agencies (Switzerland); MST (Taipei); MHESI and NSTDA (Thailand); TUBITAK and TENMAK (Turkey); NASU (Ukraine); STFC (United Kingdom); DOE and NSF (USA).
\end{acknowledgments}
\bibliography{auto_generated} 

\clearpage
\appendix
\section{Mapping between \texorpdfstring{\nch}{Nch}  and \texorpdfstring{\noff}{Ntrkoffline} in \texorpdfstring{\pp}{pp}, \texorpdfstring{\pPb}{pPb}, and \texorpdfstring{\PbPb}{PbPb} collisions}
\label{appendix:table}

\begin{table*}[h]
\centering
\topcaption{\label{tab:nch_ntrkoffline} Average multiplicity of reconstructed tracks per \nchrec bin for \nch in this analysis and \noff in previous CMS measurements~\cite{Khachatryan:2010gv, CMS:2012qk, CMS:2013jlh} in \pp, \pPb, and peripheral \PbPb collisions. Uncertainties in the tracking efficiency corrected \nch are included.}
\begin{scotch}{ l  l  l {c}@{\hspace*{5pt}} l   l {c}@{\hspace*{5pt}} l   l }
\multicolumn{1}{l}{} & \multicolumn{2}{c}{ pp\ }  && \multicolumn{2}{c}{ pPb\ } &&                   \multicolumn{2}{c}{ PbPb\ }\\
\cline{2-3}\cline{5-6}\cline{8-9}
\nchrec range & $\left<\nch\right>$ & $\left<\noff\right>$ & & $\left<\nch\right>$ &                  $\left<\noff\right>$   && $\left<\nch\right>$ & $\left<\noff\right>$ \\
\hline
$[0, 20)$    & 8$\pm$0   & 9   && 11$\pm$0  & 12    && 16$\pm$1     & 14 \\
$[20, 40)$   & 34$\pm$1  & 34   && 36$\pm$1  & 36    && 57$\pm$2     & 48 \\
$[40, 60)$   & 58$\pm$2  & 56   && 60$\pm$2  & 60    && 96$\pm$4     & 80 \\
$[60, 80)$   & 82$\pm$3  & 78   && 83$\pm$3  & 82    && 135$\pm$5     & 112 \\
$[80, 100)$   & 106$\pm$4  & 101   && 107$\pm$4  & 105    && 175$\pm$7     & 144 \\
$[100, 150)$  & 132$\pm$5  & 125   && 140$\pm$6  & 137    && 240$\pm$10     &          197 \\
$[150, 200)$   &   &                      && 198$\pm$8  & 191    && 335$\pm$13     &    276 \\
$[200, 250)$   &   &                      && 256$\pm$10  & 246    && 434$\pm$17     &    353 \\
\end{scotch}
\end{table*}

\cleardoublepage \section{The CMS Collaboration \label{app:collab}}\begin{sloppypar}\hyphenpenalty=5000\widowpenalty=500\clubpenalty=5000\input{HIN-21-012-public-authorlist.tex}\end{sloppypar}
\end{document}

%% file: HIN-21-012-public-authorlist.tex
\cmsinstitute{Yerevan Physics Institute, Yerevan, Armenia}
{\tolerance=6000
A.~Tumasyan\cmsorcid{0009-0000-0684-6742}
\par}
\cmsinstitute{Institut f\"{u}r Hochenergiephysik, Vienna, Austria}
{\tolerance=6000
W.~Adam\cmsorcid{0000-0001-9099-4341}, J.W.~Andrejkovic, T.~Bergauer\cmsorcid{0000-0002-5786-0293}, S.~Chatterjee\cmsorcid{0000-0003-2660-0349}, K.~Damanakis\cmsorcid{0000-0001-5389-2872}, M.~Dragicevic\cmsorcid{0000-0003-1967-6783}, A.~Escalante~Del~Valle\cmsorcid{0000-0002-9702-6359}, P.S.~Hussain\cmsorcid{0000-0002-4825-5278}, M.~Jeitler\cmsAuthorMark{1}\cmsorcid{0000-0002-5141-9560}, N.~Krammer\cmsorcid{0000-0002-0548-0985}, L.~Lechner\cmsorcid{0000-0002-3065-1141}, D.~Liko\cmsorcid{0000-0002-3380-473X}, I.~Mikulec\cmsorcid{0000-0003-0385-2746}, P.~Paulitsch, F.M.~Pitters, J.~Schieck\cmsAuthorMark{1}\cmsorcid{0000-0002-1058-8093}, R.~Sch\"{o}fbeck\cmsorcid{0000-0002-2332-8784}, D.~Schwarz\cmsorcid{0000-0002-3821-7331}, M.~Sonawane\cmsorcid{0000-0003-0510-7010}, S.~Templ\cmsorcid{0000-0003-3137-5692}, W.~Waltenberger\cmsorcid{0000-0002-6215-7228}, C.-E.~Wulz\cmsAuthorMark{1}\cmsorcid{0000-0001-9226-5812}
% ...
\par}
\vskip\cmsinstskip
\dag:~Deceased\\
$^{1}$Also at TU Wien, Vienna, Austria\\
% ...
